\documentstyle[12pt]{book}
\begin{document}

\sloppy
\raggedbottom

\begin{titlepage}

\hfill\vspace{1in}\\
{\Huge\bf\hspace{-\parindent}Exhibiting Randomness in \vspace{1mm}\\
Arithmetic using \vspace{1mm}\\
Mathematica and C \vspace{1.5in} \\
}
{\Large\bf G J Chaitin \vspace{1mm} \\
IBM, P O Box 704 \\
Yorktown Heights, NY 10598 \vspace{1mm} \\
{\it chaitin@watson.ibm.com} \vspace{1.5in} \\
June 6, 1993
}

\end{titlepage}

\begin{titlepage}

\hfill\\

\end{titlepage}

\markboth
{Exhibiting Randomness in Arithmetic using Mathematica \& C}{User Guide}

\chapter*{Preface / User Guide}

In my book {\it Algorithmic Information Theory\/} I explain how I
constructed a million-character equation that proves that there is
randomness in arithmetic.  My book only includes a few pages from the
monster equation, and omits the software used to construct it.  This
software has now been rewritten in {\sl Mathematica}.

The {\sl Mathematica} software for my book, and its input, are here in
their entirety.  The {\sl Mathematica} code is remarkably compact, but
it sometimes is slow.  So one {\sl C} program plus equipment for
automatically generating another is also included in this software
package.

I used Version 2.1 of {\sl Mathematica} as described in the second
edition of Wolfram's book {\it Mathematica---A System for Doing
Mathematics by Computer,} running on an {\sl IBM RISC} System/6000
workstation.

Since the {\sl APL2} character set is not generally available, I
decided to change the symbols that denote the primitive functions in
the toy {\sl LISP} that I use in {\it Algorithmic Information Theory.}

There are seven different kinds of files:

\begin{itemize}
\item Included in this distribution:

\begin{enumerate}
\item {\tt *.m} files are {\sl Mathematica} code.
\item {\tt *.c} files are {\sl C} code.
\item {\tt *.lisp} files are toy {\sl LISP} code.
      These are the four {\sl LISP} programs in my book
      ({\tt eval.lisp, eval2.lisp, eval3.lisp,} and {\tt omega.lisp}),
      plus {\tt test.lisp}.
\item {\tt *.rm} are register machine code.
\end{enumerate}

\item These will produce:

\begin{itemize}
\item[\rm 5.] {\tt *.xrm} files are expanded register machine code
      (lower level code than that in {\tt *.rm} files).
\item[\rm 6.] {\tt *.run, *.2run, *.srun, *.mrun, *.crun, *.cmrun} files
      are the output from {\sl LISP} interpreter runs.
\item[\rm 7.] {\tt *.eq} files are exponential diophantine equations.
\end{itemize}

\end{itemize}

Six different {\sl LISP} interpreters are included here:

\begin{enumerate}

\item
{\tt lisp.m} is a {\sl LISP} interpreter written in nonprocedural {\sl
Mathematica} that uses {\sl Mathematica} list structures to represent
{\sl LISP} S-expressions.  Bindings are kept in a fast look-up table.
{\tt lisp.m} converts an {\tt X.lisp} input file into an {\tt X.run}
output file.

\[
\mbox{\tt X.lisp}
\longrightarrow \fbox{\tt lisp.m} \longrightarrow
\mbox{\tt X.run}
\]

\item
{\tt lisp2.m} is a {\sl LISP} interpreter written in procedural {\sl
Mathematica} that uses {\sl Mathematica} list structures to
represent {\sl LISP} S-expressions.  Bindings are kept in a fast
look-up table.  {\tt lisp2.m} converts an {\tt X.lisp} input file into
an {\tt X.2run} output file.

\[
\mbox{\tt X.lisp}
\longrightarrow \fbox{\tt lisp2.m} \longrightarrow
\mbox{\tt X.2run}
\]

\item
{\tt slisp.m} is a {\sl LISP} interpreter written in procedural {\sl
Mathematica} that uses {\sl Mathematica} character strings to
represent {\sl LISP} S-expressions.  Bindings are kept in an
association list that must be searched sequentially.  {\tt slisp.m}
converts an {\tt X.lisp} input file into an {\tt X.srun} output file.

\[
\mbox{\tt X.lisp}
\longrightarrow \fbox{\tt slisp.m} \longrightarrow
\mbox{\tt X.srun}
\]

\item
{\tt lispm.m} is a {\sl Mathematica} program that simulates a {\sl
LISP} interpreter running on a register machine.  {\tt lispm.m}
converts an {\tt X.lisp} input file into an {\tt X.mrun} output file.

Before running this program, {\tt xpnd.m} must be used to convert
{\tt lisp.rm} into {\tt lisp.xrm}, which is needed by this program.

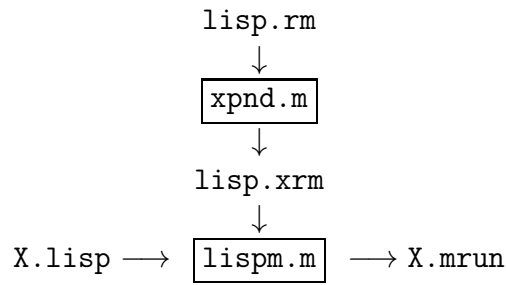
\[
\mbox{\tt X.lisp}
\longrightarrow
\begin{array}[b]{c}
\mbox{\tt lisp.rm} \\
\downarrow         \\
\fbox{\tt xpnd.m}  \\
\downarrow         \\
\mbox{\tt lisp.xrm}\\
\downarrow         \\
\fbox{\tt lispm.m}
\end{array}
\longrightarrow
\mbox{\tt X.mrun}
\]

\item
{\tt clisp.m} is a {\sl Mathematica} program serving as a driver for a
{\sl LISP} interpreter written in {\sl C}.  {\tt clisp.m} converts an
{\tt X.lisp} input file into an {\tt X.crun} output file.

Before running {\tt clisp.m}, the {\sl C} program {\tt lisp.c} must be
compiled using the command {\tt cc -O -olisp lisp.c}.

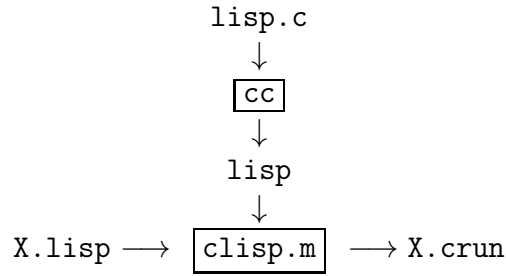
\[
\mbox{\tt X.lisp}
\longrightarrow
\begin{array}[b]{c}
\mbox{\tt lisp.c}\\
\downarrow       \\
\fbox{\tt cc}    \\
\downarrow       \\
\mbox{\tt lisp}  \\
\downarrow       \\
\fbox{\tt clisp.m}
\end{array}
\longrightarrow
\mbox{\tt X.crun}
\]

\item
{\tt clispm.m} is a {\sl Mathematica} program serving as a driver for
a {\sl C} program that simulates a {\sl LISP} interpreter running on a
register machine.  {\tt clispm.m} converts an {\tt X.lisp} input file
into an {\tt X.cmrun} output file.

Before running {\tt clispm.m}, {\tt xpnd.m} must be used to convert
{\tt lisp.rm} into {\tt lisp.xrm}.  {\tt rm2c.m} must then be used to
convert {\tt lisp.xrm} into the {\sl C} program {\tt lispm.c}.  {\tt
lispm.c} is then compiled using the command {\tt cc -O -olispm
lispm.c}.

\[
\mbox{\tt X.lisp}
\longrightarrow
\begin{array}[b]{c}
\mbox{\tt lisp.rm} \\
\downarrow         \\
\fbox{\tt xpnd.m}  \\
\downarrow         \\
\mbox{\tt lisp.xrm}\\
\downarrow         \\
\fbox{\tt rm2c.m}  \\
\downarrow         \\
\mbox{\tt lispm.c} \\
\downarrow         \\
\fbox{\tt cc}      \\
\downarrow         \\
\mbox{\tt lispm}   \\
\downarrow         \\
\fbox{\tt clispm.m}
\end{array}
\longrightarrow
\mbox{\tt X.cmrun}
\]

\end{enumerate}

To run any one {\tt X.m} of these six {\sl LISP} interpreters, first
enter {\sl Mathematica} using the command {\tt math}.  Then tell
{\sl Mathematica},
\[
\mbox{\tt << X.m}
\]
To run a {\sl LISP} program {\tt X.lisp}, enter
\[
\mbox{\tt run @ "X"}
\]
To run several programs, enter
\begin{center}
\verb!run /@ {"X","Y","Z"}!
\end{center}
Before changing to another {\sl LISP} interpreter, type {\tt Exit} to
exit from {\sl Mathematica}, and then begin a fresh {\sl Mathematica}
session.

Here is how to run the {\sl LISP} test program, the three {\sl LISP} in {\sl
LISP} examples in my book, and then start computing the halting
probability $\Omega$ in the limit from below:

\begin{verbatim}
               math
               << clispm.m
               run @ "test"
               run /@ {"eval","eval2","eval3"}
               Exit

               math
               << clisp.m
               run @ "omega"
               Exit
\end{verbatim}

The six different {\sl LISP} interpreters run at vastly different
speeds, but should always produce identical results.  This can easily
be checked, for example, as follows:

\begin{verbatim}
               diff X.run X.crun > out
               vi out
\end{verbatim}

Two different front ends are available for these six {\sl LISP}
interpreters:

\begin{enumerate}

\item
{\tt run.m} is written in procedural {\sl Mathematica}.  As each
M-expression is read in, it is written out, then converted
to an S-expression that is written out and evaluated.\footnote{The
conversion from M- to S-expression mostly consists of making all
implicit parentheses explicit.}

\item
{\tt run2.m} is written in non-procedural {\sl Mathematica}.  All
M-expressions are read in at once.  Then each is converted to an
S-expression that is written out and evaluated.

\end{enumerate}

Which front end is used is determined by {\tt frontend.m}.  Each of
the six {\sl LISP} interpreters contains a {\tt <<} of {\tt
frontend.m}.  Normally {\tt frontend.m} is {\tt << run.m} and the
first front end is chosen.  To select the second front end, change
this to {\tt << run2.m}.

\[
\fbox{{\sl LISP interpreter}{\tt .m}}
\mbox{\tt <<}
\fbox{\tt frontend.m}
\begin{array}{l}
\mbox{\tt <<} \fbox{\tt run.m} \\
\mbox{\tt <<} \fbox{\tt run2.m} \\
\end{array}
\]

Three register machine programs {\tt *.rm} are provided: {\tt
example.rm}, {\tt test.rm}, and {\tt lisp.rm}.  {\tt example.rm} is
the tiny example given in my Cambridge book.  {\tt test.rm} has each
possible register machine instruction, but it is not a program that
can be run.  {\tt lisp.rm} is the {\sl LISP} interpreter used by {\tt
lispm.m} and {\tt clispm.m}, and converted into the monster
exponential diophantine equation by {\tt eq.m}.

More precisely, to convert any one of the three register machine
programs {\tt X.rm} into an exponential diophantine equation there are
two steps.  First, use {xpnd.m} to convert {\tt X.rm} into {\tt
X.xrm}.  Then use {\tt eq.m} to convert {\tt X.xrm} into {\tt X.eq}.
For more output, set {\tt fulloutput = True} before typing {\tt <<
eq.m}.  For each conversion, a fresh copy of {\tt eq.m} must be loaded
into a clean {\sl Mathematica} session.

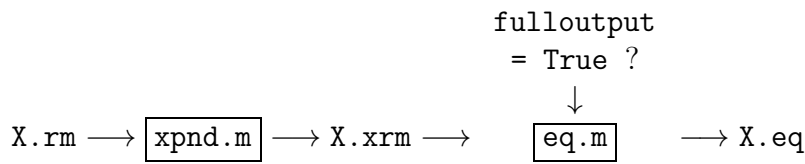
\[
\mbox{\tt X.rm} \longrightarrow
\fbox{\tt xpnd.m} \longrightarrow
\mbox{\tt X.xrm} \longrightarrow
\begin{array}[b]{c}
\mbox{\tt fulloutput} \\
\mbox{\tt = True \rm ?} \\
\downarrow \\
\fbox{\tt eq.m}
\end{array}
\longrightarrow
\mbox{\tt X.eq}
\]

Here is how to generate the monster equation:

\begin{verbatim}
               math
               << xpnd.m
               run @ "lisp"
               Exit

               math
              [fulloutput = True]
               << eq.m
               fn of fn.xrm file = lisp
               Exit
\end{verbatim}

How does this software help to exhibit randomness in arithmetic?

Take the equation in {\tt lisp.eq}.  Substitute 0 for {\tt
input[reg\$X]} for each register {\tt reg\$X} except for {\tt
reg\$expression}.  Substitute a toy {\tt LISP} expression that halts if
and only if (the $k$th bit of the $n$th approximation to $\Omega$ is
1) for {\tt input[reg\$expression]}.  (Most of the pieces for this are
in {\tt omega.lisp}.)  The resulting exponential diophantine
equation is $1.
\times 10^6$ characters long and has $2. \times 10^4$ variables.  It
has exactly one solution for a given value of $k$ and $n$ if the $k$th
bit of the $n$th approximation to $\Omega$ is 1.  It has no solutions
for a given value of $k$ and $n$ if the $k$th bit of the $n$th
approximation to $\Omega$ is 0.  Now think of $n$ as a variable rather
than as a parameter.  The resulting equation has only finitely many
solutions if the $k$th bit of $\Omega$ is 0.  It has infinitely
many solutions if the $k$th bit of $\Omega$ is 1.

\section*{Bibliography}

\begin{itemize}
\item[{[1]}] S. Wolfram, {\it Mathematica---A System for Doing
Mathematics by Computer,} second edition, Addison-Wesley, 1991.
\item[{[2]}] B. W. Kernighan and D. M. Ritchie, {\it The C Programming
Language,} second edition, Prentice Hall, 1988.
\item[{[3]}] G. J. Chaitin, {\it Algorithmic Information Theory,}
revised third printing, Cambridge University Press, 1990.
\item[{[4]}] G. J. Chaitin, {\it Information, Randomness \&
Incompleteness,} second edition, World Scientific, 1990.
\item[{[5]}] G. J. Chaitin, {\it Information-Theoretic
Incompleteness,} World Scientific, 1992.
\item[{[6]}] G. J. Chaitin, ``Randomness in arithmetic and the decline
and fall of reductionism in pure mathematics,'' {\it IBM Research
Report RC-18532,} November 1992.
\end{itemize}

\tableofcontents

\newcommand
{\chap}[1]{\chapter*{#1}\markright{#1}\addcontentsline{toc}{chapter}{#1}}
\newcommand
{\size}{\small}   

{
}\chap{eq.m}{\size\begin{verbatim}
(***** EQ.M *****)

fulloutput = If[ fulloutput, True, False, False ]
fn = InputString["fn of fn.xrm file = "]
t0 = SessionTime[]
p = Get[fn<>".xrm"] (* read in program *)
o = OpenWrite[fn<>".eq",PageWidth->62]
Format[LineBreak[_]] = ""
Format[Continuation[_]] = " "
print[x_] := (Print@ x; Write[o,OutputForm@ x])

print@
 "********** program"
print@
 Short[InputForm@ p,10]

(* get set of labels of all instructions in program *)

labels = #[[1]]& /@ p

If[
 Length@ Union@ labels != Length@ p,
 print@
 "Duplicate labels!"
]

(* get set of all registers in program *)

registers = Union@ Flatten@ (Drop[#,2]& /@ p)
registers = Cases[registers,_Symbol]
registers = Complement[registers,labels]

eqs = {}
put[x_] := (Write[o,x]; eqs = {eqs,x};)
Write[o,OutputForm@
 "********** <='s & =='s as they are generated"
 ]

{
 (* generate equations for base q *)
 totalinput == Plus@@ (input[#]& /@ registers),
 numberofinstructions == Length@ p,
 longestlabel == (* with ( ) around label for jump's *)
 Max@ (StringLength["("<>ToString[#]<>")"]& /@ labels),
 q == 256^
 (totalinput+ time+ numberofinstructions+ longestlabel+ 1),
 qminus1 + 1 == q,
 1 + q i == i + q^time,
 (* label equations *)
 (# <= i)& /@ labels,
 i == Plus@@ labels,
 (* equations for starting & halting *)
 1 <= p[[1,1]],
 q^time == q Plus@@ Cases[p,{l_,halt}->l]
} // put

(* generate flow equations *)

Evaluate[ next /@ labels ] = RotateLeft@ labels

{
 Cases[ p, {l_,goto,l2_} -> q l <= l2 ],
 Cases[ p, {l_,jump,a_,l2_} -> q l <= l2 ],
 Cases[ p, {l_,goback,a_} ->
 (
 { goback <= x,
 goback <= qminus1 l,
 x <= goback + qminus1 (i-l)
 } /.
 goback -> goback[l] /.
 { {x -> a}, {x -> nextic} }
 )
 ],
 Cases[ p, {l_,eq|eqi,a_,b_,l2_} ->
 {
 q l <= next[l] + l2,
 q l <= next[l] + q eq[a,b]
 }
 ],
 Cases[ p, {l_,neq|neqi,a_,b_,l2_} ->
 {
 q l <= next[l] + l2,
 q l <= l2 + q eq[a,b]
 }
 ],
 Cases[
 DeleteCases[ p,
 {_,halt|goto|jump|goback|eq|eqi|neq|neqi,___}
 ],
 {l_,__} -> q l <= next[l]
 ],
 {
 ic == Plus@@ ((# "("<>ToString[#]<>")")& /@ labels),
 q nextic <= ic,
 ic <= q nextic + qminus1
 }
} // put

(* generate compare equations *)

(
 Cases[ p, {l_,eq|neq,a_,b_,_} ->
 compare[a,b,char[a],char[b]]
 ]
 ~Union~
 Cases[ p, {l_,eqi|neqi,a_,b_,_} ->
 compare[a,b,char[a],b i]
 ]
) /.
 compare[a_,b_,charA_,charB_] ->
{
 {
 eq[a,b] <= i,
 2 eq[a,b] <= ge[a,b] + ge[b,a],
 ge[a,b] + ge[b,a] <= 2 eq[a,b] + i
 },
 {
 geXY <= i,
 256 geXY <= 256 i + charX - charY,
 256 i + charX - charY <= 256 geXY + 255 i
 } /.
 {
 {geXY -> ge[a,b], charX -> charA, charY -> charB},
 {geXY -> ge[b,a], charX -> charB, charY -> charA}
 }
} // put

(* generate auxiliary register equations *)

(* set target t to source s at label l *)
set[t_,s_,l_] :=
 {
 set <= s,
 set <= qminus1 l,
 s <= set + qminus1 (i - l)
 } /.
 set -> set[t,l]

{
 Cases[ p, {l_,set,a_,b_} ->
 set[a,b,l]
 ],
 Cases[ p, {l_,seti,a_,b_} ->
 set[a,b i,l]
 ],
 Cases[ p, {l_,left,a_,b_} ->
 {
 set[a,256a+char[b],l],
 set[b,shift[b],l]
 }
 ],
 Cases[ p, {l_,lefti,a_,b_} ->
 set[a,256a+i b,l]
 ],
 Cases[ p, {l_,right,a_} ->
 set[a,shift[a],l]
 ],
 Cases[ p, {l_,jump,a_,_} :>
 set[a,i "("<>ToString[next[l]]<>")",l]
 ]
} // put

(* generate main register equations *)

defs[r_] := defs[r] = Cases[ p,
 {l_,set|seti|left|lefti|right|jump,r,___} |
 {l_,left,_,r}
 -> l ]

(
 Function[ r,
 {
 r <= qminus1 i,
 r + output q^time ==
 input + q (dontset + Plus@@ (set2[r,#]& /@ defs[r])),
 set == Plus@@ defs[r],
 dontset <= r,
 dontset <= qminus1 (i - set),
 r <= dontset + qminus1 set,
 256 shift <= r,
 256 shift <= i (qminus1 - 255),
 r <= 256 shift + 255 i,
 r == 256 shift + char
 } /. ((# -> #[r])& /@
 {input,output,set,dontset,shift,char}) /.
 set2 -> set
 ] /@ registers
) // put

(* all equations and inequalities are now in eqs; *)
(* start processing *)

eqs = Flatten[eqs]

print@
 "********** combined list of <='s & =='s"
print@
 Short[InputForm@ eqs,10]

(* how many ='s, <='s, registers, labels, variables ? *)

print@StringForm[
 "********** `` =='s, `` <='s, `` total",
 neq = Count[eqs,_==_], nle = Count[eqs,_<=_], Length@ eqs
 ]
print@
 "********** now counting variables"

variables =
 eqs /. Plus|Times|Power|Equal|LessEqual -> List

variables =
 DeleteCases[ Flatten@ variables, _String|_Integer ] // Union

print@StringForm[
"********** `` registers, `` labels, `` variables altogether",
Length@ registers, Length@ labels, nvar = Length@ variables
]
Write[o,variables]

(* convert strings to integers *)

alphabet = "\000()" ~StringJoin~
 "ABCDEFGHIJKLMNOPQRSTUVWXYZ" ~StringJoin~
 "abcdefghijklmnopqrstuvwxyz" ~StringJoin~
 "0123456789_+-.',!=*&?/:\"$"

bitmap =
 MapThread[
 #1 -> StringJoin[
 Rest@ IntegerDigits[256 + #2, 2] /.
 {0 -> "0", 1 -> "1"}
 ] & ,
 { Characters@ alphabet, Range[0, StringLength@ alphabet -1] }
 ]

s2i[x_] :=
 ToExpression[
 "2^^" <> StringReverse@ StringReplace[x,bitmap]
 ]

print@
 "********** now converting strings to integers"

eqs = eqs /.
 {eq[x__] -> eq[x], ge[x__] -> ge[x], x_String :> s2i@x}

(* transpose negative terms from rhs to lhs of equation *)

negterms[ (term:(x_Integer _. /; x < 0)) + rest_. ] :=
 term + negterms@ rest

negterms[ _ ] := 0

fix[x_] :=
 (
 x /. l_ == r_ :> l == Expand @ r
 ) /. l_ == r_ :> ( (l - # == r - #)&@ negterms@ r )

(* expand each implication into 7 equations & *)
(* add 9 variables *)

print@
 "********** now expanding <='s"
If[ fulloutput,
 Write[o,OutputForm@
 "********** expand each <="
 ]
]

eqs = eqs /. a_ <= b_ :>
(
 If[ fulloutput, Write[o,a<=b]; Write[o,#]; #, # ]& @
 Module[ {r,s,t,u,v,w,x,y,z},
 {
 fix[r == a],
 fix[s == b],
 t == 2^s,
 (1+t)^s == v t^(r+1) + u t^r + w,
 w + x + 1 == t^r,
 u + y + 1 == t,
 u == 2 z + 1
 }
 ]
)

eqs = Flatten[eqs]

print@
 "********** <='s expanded into =='s"
print@
 Short[InputForm@ eqs,10]
print@
 "********** each <= became 7 =='s and added 9 variables"
print@StringForm[
 "********** so should now have `` =='s and `` variables",
 neq + 7 nle, nvar + 9 nle
 ]
print@StringForm[
 "********** actually there are now `` =='s",
 Length@ eqs
 ]

(* combine all equations into one equation *)

ClearAttributes[ {Plus,Times}, {Orderless,Flat} ]

print@
"********** now combining equations"

eqn =
(
 Plus@@ ( eqs /. l_ == r_ -> (l^2 + r^2) ) ==
 Plus@@ ( eqs /. l_ == r_ -> 2 l r )
)

(***
(* Check that no =='s or <='s have become True or False, *)
(* that no <='s are left, that there are no minus signs, *)
(* and that there is just one == *)
If[ fulloutput,
 trouble[] := (Print@"trouble!"; Abort[]);
 print@
 "********** now checking combined equation";
 eqn /. True :> trouble[];
 eqn /. False :> trouble[];
 eqn /. _<=_ :> trouble[];
 eqn /. x_Integer /; x < 0 :> trouble[];
 eqn[[1]] /. _==_ :> trouble[];
 eqn[[2]] /. _==_ :> trouble[];
]
***)

print@
"********** combined equation"
print@
 Short[InputForm@ eqn,10]
print@StringForm[
 "********** `` terms on lhs, `` terms on rhs",
 Length@ eqn[[1]], Length@ eqn[[2]]
 ]
Write[o,OutputForm@
 "********** combined equation 2"
 ]
Write[o,OutputForm@
 Short[InputForm@ eqn,100]
 ]
Write[o,OutputForm@
 "********** left side"
 ]
Write[o,OutputForm@
 Short[InputForm@ eqn[[1]],50]
 ]
Write[o,OutputForm@
 "********** right side"
 ]
Write[o,OutputForm@
 Short[InputForm@ eqn[[2]],50]
 ]
Write[o,OutputForm@
 "********** first 50 terms"
 ]
Write[o,
 Take[eqn[[1]],+50]
 ]
Write[o,OutputForm@
 "********** last 50 terms"
 ]
Write[o,
 Take[eqn[[2]],-50]]
If[ fulloutput,
 print@
 "********** now writing full equation";
 Write[o,OutputForm@
 "********** combined equation in full"
 ];
 Write[o,
 eqn
 ],
 print@
 "********** now determining size of equation";
 print@StringForm[
 "********** size of equation `` characters",
 StringLength@ ToString@ InputForm@ eqn
 ]
]
print@StringForm[
 "********** elapsed time `` seconds",
 Round[SessionTime[]-t0]
 ]
Print@
 "********** list of =='s left in variable eqs"
Print@
 "********** combined == left in variable eqn"
Print@
 "********** warning: + * noncommutative nonassociative!"
Print@
 "********** (to preserve order of terms & factors in eqn)"
Close@ o
\end{verbatim}
}\chap{lisp.m}{\size\begin{verbatim}
(***** LISP.M *****)

<<frontend.m

(* "nonprocedural" lisp interpreter *)

identitymap =
 ( FromCharacterCode /@ Range[0,255] ) ~Join~ {{},}

pos[c_String] :=
 ( If[ # <= 256, #, Abort[] ] )& @
 ( 1 + First@ ToCharacterCode@ c )
pos[{}] :=
 257
pos[_] :=
 258

eval[e_,,d_] :=
 eval[e,identitymap,d]

eval[(e:({}|_String)),a_,_] :=
 a[[ pos@ e ]]

eval[e_,a_,d_] :=
 eval[ eval[ First@ e,a,d ], Rest@ e, a, d ]

eval["'",{e_:{},___},_,_] :=
 e

eval["/",{p_:{},q_:{},r_:{},___},a_,d_] :=
If[
 eval[p,a,d] =!= "0",
 eval[q,a,d],
 eval[r,a,d]
]

eval[f_,e_,a_,d_] :=
 apply[ f, eval[#,a,d]& /@ e, a, d ]

apply["+",{},_,_] := {}
apply["+",{{},___},_,_] := {}
apply["+",{x_String,___},_,_] := x
apply["+",{x_,___},_,_] := First@ x

apply["-",{},_,_] := {}
apply["-",{{},___},_,_] := {}
apply["-",{x_String,___},_,_] := x
apply["-",{x_,___},_,_] := Rest@ x

apply["*",{x_,_String,___},_,_] := x
apply["*",{x_:{},y_:{},___},_,_] := {x} ~Join~ y

apply[".",{},_,_] := "1"
apply[".",{{},___},_,_] := "1"
apply[".",{_String,___},_,_] := "1"
apply[".",_,_,_] := "0"

apply["=",{x_:{},y_:{},___},_,_] :=
 If[ x === y, "1", "0" ]

apply[",",{x_:{},___},_,_] :=
 (print[ "display", output@ x ]; x)

apply["!",_,_,d_] :=
 Throw@ "?" /; d == 0
apply["!",{x_:{},___},_,d_] :=
 eval[x,,d-1]

apply["?",_,_,d_] :=
 Throw@ "?" /; d == 0
apply["?",{}|{_},_,_] :=
 {}
apply["?",{_String,y_,___},_,d_] :=
 apply["?",{{},y},,d]
apply["?",{x_,y_,___},_,d_] :=
 Catch@ {eval[y,,Length@x]} /; Length@x < d-1
apply["?",{x_,y_,___},_,d_] :=
 Catch@ {eval[y,,d-1]} // If[ # === "?", Throw@ #, # ] &

(* If not a primitive function: *)
apply[_,_,_,d_] :=
 Throw@ "?" /; d == 0
apply[(b:({}|_String)),_,a_,_] :=
 a[[ pos@ b ]]
apply[{_,_String,b_:{},___},_,a_,d_] :=
 eval[b,a,d-1]
apply[{_,x_:{},b_:{},___},v_,a_,d_] :=
 eval[ b, bind[x,v,a], d-1 ]

bind[{},v_,a_] :=
 a

bind[x_,{},a_] :=
 bind[x,{{}},a]

bind[x_,v_,a_] :=
ReplacePart[
 bind[ Rest@ x, Rest@ v, a ],
 First@ v,
 pos@ First@ x
]

eval[e_] :=
(
 print[ "expression", output@ e ];
 eval[ wrap@ e,,Infinity ]
)

run[fn_] := run[fn, "lisp.m", ".run"]
\end{verbatim}
}\chap{lisp2.m}{\size\begin{verbatim}
(***** LISP2.M *****)

<<frontend.m

(* "procedural" lisp interpreter *)

identitymap =
 ( FromCharacterCode /@ Range[0,255] ) ~Join~ {{},}

pos[c_String] :=
 ( If[ # <= 256, #, Abort[] ] )& @
 ( 1 + First@ ToCharacterCode@ c )
pos[{}] :=
 257
pos[_] :=
 258

at[x_] :=
 MatchQ[ x, {}|_String ]
hd[x_] :=
 If[ at@ x, x, First@ x ]
tl[x_] :=
 If[ at@ x, x, Rest@ x ]
jn[x_,y_] :=
 If[ MatchQ[y,_String], x, Prepend[y,x] ]

eval[e_,,d_] := eval[e,identitymap,d]

eval[e2_,a_,d2_] :=

Block[ {e = e2, d =d2, f, args, x, y},
 If[ at@ e, Return@ a[[ pos@ e ]] ];
 f = eval[hd@ e,a,d];
 e = tl@ e;
 Switch[
 f,
 "'", Return@ hd@ e,
 "/", Return@
 If[
 eval[hd@ e,a,d] =!= "0",
 eval[hd@tl@ e,a,d],
 eval[hd@tl@tl@ e,a,d]
 ]
 ];
 args = eval[#,a,d]& /@ e;
 x = hd@ args;
 y = hd@tl@ args;
 Switch[
 f,
 "+", Return@ hd@ x,
 "-", Return@ tl@ x,
 "*", Return@ jn[x,y],
 ".", Return@ If[ at@ x, "1", "0" ],
 "=", Return@ If[ x === y, "1", "0" ],
 ",", Return@ (print[ "display", output@ x ]; x)
 ];
 If[ d == 0, Throw@ "?" ];
 d--;
 Switch[
 f,
 "!", Return@ eval[x,,d],
 "?", Return@
 If[
 Length@x < d,
 Catch@ {eval[y,,Length@x]},
 Catch@ {eval[y,,d]} //
 If[ # === "?", Throw@ #, # ] &
 ]
 ];
 f = tl@ f;
 eval[ hd@tl@ f, bind[hd@ f,args,a], d ]
]

bind[vars_?at,args_,a_] :=
 a

bind[vars_,args_,a_] :=
ReplacePart[
 bind[ tl@ vars, tl@ args, a ],
 hd@ args,
 pos@ hd@ vars
]

eval[e_] :=
(
 print[ "expression", output@ e ];
 eval[ wrap@ e,,Infinity ]
)

run[fn_] := run[fn, "lisp2.m", ".2run"]
\end{verbatim}
}\chap{slisp.m}{\size\begin{verbatim}
(***** SLISP.M *****)

<<frontend.m

(* string lisp interpreter *)

at[x_] := StringLength@ x == 1 || x === "()"

hd[x_] :=
(If[ at@ x, Return@ x ];
 Block[ {p = 0},
 Do[
 p += Switch[ StringTake[x,{i}], "(", +1, ")", -1, _, 0 ];
 If[ p == 0, Return@ StringTake[x,{2,i}] ],
 {i, 2, StringLength@ x}
 ]
 ]
)

tl[x_] :=
(If[ at@ x, Return@ x ];
 Block[ {p = 0},
 Do[
 p += Switch[ StringTake[x,{i}], "(", +1, ")", -1, _, 0 ];
 If[ p == 0, Return[ "("<>StringDrop[x,i] ] ],
 {i, 2, StringLength@ x}
 ]
 ]
)

jn[x_,y_] :=
 If[ StringLength@ y == 1, x, "("<>x<>StringDrop[y,1] ]

eval[e_,,d_] := eval[e,"()",d]

eval[e2_,a_,d2_] :=

Block[ {e = e2, d = d2, f, args, x, y},
 If[
 at@ e,
 Return@
 Which[
 e === hd@ a, hd@tl@ a,
 at@ a, e,
 True, eval[ e, tl@tl@ a, ]
 ]
 ];
 f = eval[ hd@ e, a, d ];
 e = tl@ e;
 Switch[
 f,
 "'", Return@ hd@ e,
 "/", Return@
 If[
 eval[hd@ e,a,d] =!= "0",
 eval[hd@tl@ e,a,d],
 eval[hd@tl@tl@ e,a,d]
 ]
 ];
 args = evlst[e,a,d];
 x = hd@ args;
 y = hd@tl@ args;
 Switch[
 f,
 "+", Return@ hd@ x,
 "-", Return@ tl@ x,
 "*", Return@ jn[x,y],
 ".", Return@ If[ at@ x, "1", "0" ],
 "=", Return@ If[ x === y, "1", "0" ],
 ",", Return@ (print[ "display", output@ x ]; x)
 ];
 If[ d == 0, Throw@ "?" ];
 d--;
 Switch[
 f,
 "!", Return@ eval[x,,d],
 "?", Return@
 If[ size@x < d,
 Catch[ "("<>eval[y,,size@x]<>")" ],
 Catch[ "("<>eval[y,,d]<>")" ] //
 If[ # === "?", Throw@ #, # ] &
 ]
 ];
 f = tl@ f;
 eval[ hd@tl@ f, bind[hd@ f,args,a], d ]
]

size[x_?at] := 0
size[x_] := 1 + size@ tl@ x

evlst[e_?at,a_,d_] := e
evlst[e_,a_,d_] := jn[ eval[hd@ e,a,d], evlst[tl@ e,a,d] ]

bind[vars_?at,args_,a_] := a
bind[vars_,args_,a_] :=
 jn[hd@ vars, jn[hd@ args, bind[tl@ vars,tl@ args,a]]]

eval[e_] :=
(
 print[ "expression", output@ e ];
 eval[ output@ wrap@ e,,Infinity ]
)

run[fn_] := run[fn, "slisp.m", ".srun"]
\end{verbatim}
}\chap{lispm.m}{\size\begin{verbatim}
(***** LISPM.M *****)

<<frontend.m

(* lisp machine interpreter *)

p = << lisp.xrm

labels = Cases[p, {l_,__} -> l]

If[
 Length@ Union@ labels != Length@ p,
 Print@ "Duplicate labels!!!"
]

registers = Cases[p, {_,_,r__} -> r] // Flatten // Union
registers = Cases[registers, r_Symbol -> r]
registers = Complement[registers,labels]

Evaluate[ next /@ labels ] = RotateLeft@ labels
Evaluate[ #[]& /@ registers ] = {}& /@ registers
Evaluate[ #[]& /@ labels ] =
 Cases[p, {l_,op_,x___} -> op[next[l],x]]

first[x_] := If[ x === {}, "\0", x[[1]] ]

out[n_,r_] :=
(
 print[ "display", StringJoin@@ Flatten@ r[] ];
 n
)

dump[n_] :=
(
 print[ ToString@ #, StringJoin@@ Flatten@ #[] ] & /@
 registers;
 n
)

eqi[n_,r_,i_,l_] := If[ first[r[]] === i, l, n ]
neqi[n_,r_,i_,l_] := If[ first[r[]] =!= i, l, n ]
eq[n_,r_,s_,l_] := If[ first[r[]] === first[s[]], l, n ]
neq[n_,r_,s_,l_] := If[ first[r[]] =!= first[s[]], l, n ]

lefti[n_,r_,i_] :=
If[
 i === "\0", error[],
 r[] = {i, r[]};
 n
]

left[n_,r_,s_] :=
If[
 s[] === {}, error[],
 r[] = {s[][[1]], r[]};
 s[] = s[][[2]];
 n
]

right[n_,r_] :=
If[
 r[] === {}, error[],
 r[] = r[][[2]];
 n
]

seti[n_,r_,"\0"] := (r[] = {}; n)
seti[n_,r_,i_] := (r[] = {i, {}}; n)
set[n_,r_,s_] := (r[] = s[]; n)

goto[n_,l_] := l
halt[n_] := halt
error[] := (Print@ "ERROR!!!"; Abort[])

ravel[c_,r___] := {c, ravel[r]}
ravel[] := {}

jump[n_,r_,l_] :=
(
 r[] = ravel@@ Characters[ "("<>ToString[n]<>")" ];
 l
)

goback[n_,r_] :=
ToExpression[
 StringJoin@@ Drop[ Drop[ Flatten@ r[], 1], -1]
]

eval[e_] :=
(
 print[ "expression", output@ e ];
 reg$expression[] = ravel@@ Characters@ output@ wrap@ e;
 loc = lab$l1;
 While[ loc =!= halt, clock++; loc = loc[] ];
 StringJoin@@ Flatten@ reg$value[]
)

run[fn_] := run[fn, "lispm.m", ".mrun"]
\end{verbatim}
}\chap{clisp.m}{\size\begin{verbatim}
(* CLISP.M *)

<<frontend.m

(* driver for C lisp interpreter *)

eval[e_] :=
(
 print[ "expression", output@ e ];
 t1 = "tmp1"<>ToString@ Random[Integer,10^10];
 t2 = "tmp2"<>ToString@ Random[Integer,10^10];
 tmp1 = OpenWrite@ t1;
 (* should check that input has no \0 characters *)
 (* and also no characters above hex FF *)
 WriteString[tmp1, output@ wrap@ e,"\n"];
 Close@ tmp1;
 Run["lisp","<",t1,">",t2];
 tmp2 = ReadList[t2,Record];
 Run["rm",t1];
 Run["rm",t2];
 print["display",#]& /@ Drop[tmp2,-1];
 tmp2[[-1]]
)

run[fn_] := run[fn, "clisp.m", ".crun" ]
\end{verbatim}
}\chap{clispm.m}{\size\begin{verbatim}
(* CLISPM.M *)

<<frontend.m

(* driver for C lisp machine *)

eval[e_] :=
(
 print[ "expression", output@ e ];
 t1 = "tmp1"<>ToString@ Random[Integer,10^10];
 t2 = "tmp2"<>ToString@ Random[Integer,10^10];
 tmp1 = OpenWrite@ t1;
 (* should check that input has no \n or \0 characters *)
 WriteString[tmp1, StringReverse@ output@ wrap@ e,"\n"];
 Close@ tmp1;
 Run["lispm","<",t1,">",t2];
 tmp2 = ReadList[t2,Record];
 Run["rm",t1];
 Run["rm",t2];
 clock = ToExpression@ tmp2[[-1]];
 tmp2 = StringReverse /@ Drop[tmp2,-1];
 print["display",#]& /@ Drop[tmp2,-1];
 tmp2[[-1]]
)

run[fn_] := run[fn, "clispm.m", ".cmrun"]
\end{verbatim}
}\chap{frontend.m}{\size\begin{verbatim}
(* FRONTEND.M *)

<<run.m

(* or <<run2.m *)
\end{verbatim}
}\chap{run.m}{\size\begin{verbatim}
(***** RUN.M *****)

(* handle {dd} chars *)
t[x_] := StringReplace[x,convertmap]
convertmap =
 ( FromCharacterCode@ # -> ToString@ {#-128} )& /@
 Range[128,255]
convertmap2 = convertmap /. (l_->r_)->(r->l)

chr3[]:=
Block[ {c},
While[
 StringLength@ line == 0,
 line = Read[i,Record];
 If[ line == EndOfFile, Abort[] ];
 Print@ line;
 WriteString[o,line,"\n"];
 (* keep only non-blank printable ASCII codes *)
 line = FromCharacterCode@
 Cases[ ToCharacterCode@ line, n_Integer /; 32 < n < 127 ]
];
c = StringTake[line,1];
line = StringDrop[line,1];
c
]

chr2[] :=
Block[ {c},
 c = chr3[];
 If[ c =!= "{", Return@ c ];
 While[ StringTake[c,-1] =!= "}", c = c<>chr3[] ];
 c = StringReplace[c,convertmap2];
 If[ StringLength@ c == 1, Return@ c ];
 StringReplace["{0}",convertmap2]
]

chr[] :=
Block[ {c},
While[ True,
 c = chr2[];
 If[ c =!= "[", Return@ c ];
 While[ chr[] =!= "]" ]
]
]

get[sexp_:False,rparenokay_:False] :=

Block[ {c = chr[], d, l ={}, name, def, body, varlist},
 Switch[
 c,
 ")", Return@ If[rparenokay,")",{}],
 "(",
 While[ ")" =!= (d = get[sexp,True]),
 AppendTo[l,d]
 ];
 Return@ l
 ];
 If[ sexp, Return@ c ];
 Switch[
 c,
 "\"", get[True],
 ":",
 {name,def,body} = {get[],get[],get[]};
 If[
 !MatchQ[name,{}|_String],
 varlist = Rest@ name;
 name = First@ name;
 def = {"'",{"&",varlist,def}}
 ];
 {{"'",{"&",{name},body}},def},
 "+"|"-"|"."|"'"|","|"!", {c,get[]},
 "*"|"="|"&"|"?", {c,get[],get[]},
 "/"|":", {c,get[],get[],get[]},
 _, c
 ]
]

(* output S-exp *)
output[x_String] := x
output[{x___}] := StringJoin["(", output /@ {x}, ")"]

blanks = StringJoin@ Table[" ",{12}]

print[x_,y_] := print1[t@ x,t@ y]
print1[x_,y_] := (print2[x,StringTake[y,50]];
 print1["",StringDrop[y,50]]) /; StringLength[y] > 50
print1[x_,y_] := print2[x,y]
print2[x_,y_] := print3[StringTake[x<>blanks,12]<>y]
print3[x_] := (Print[x]; WriteString[o,x,"\n"])

wrap[e_] :=
If[ names === {}, e, {{"'",{"&",names,e}}} ~Join~ defs ]

let[n_,d_] :=
(
 print[ output@ n<> ":", output@ d ];
 names = {n} ~Join~ names;
 defs = {{"'",d}} ~Join~ defs;
)

run[fn_,whoami_,outputsuffix_] :=
(
 line = "";
 names = defs = {};
 t0 = SessionTime[];
 o = OpenWrite[fn<>outputsuffix];
 i = OpenRead[fn<>".lisp"];
 print3["Start of "<>whoami<>" run of "<>fn<>".lisp"];
 print3@ "";
 CheckAbort[
 While[True,
(print3@ "";
 clock = 0;
 Replace[#,{
 {"&",{func_,vars___},def_} :> let[func,{"&",{vars},def}],
 {"&",var_,def_} :> let[var,eval@ def],
 _ :> print[ "value", output@ eval@ # ]
 }];
 If[clock != 0, print["cycles",ToString@clock]]
)& @ get[];
 print3@ ""
 ],
 ];
 print3@ StringForm[
 "Elapsed time `` seconds",
 Round[SessionTime[]-t0]
 ];
 Close@ i;
 Close@ o
)

runall := run /@ {"test","eval","eval2","eval3","omega"}

$RecursionLimit = $IterationLimit = Infinity
SetOptions[$Output,PageWidth->63];
\end{verbatim}
}\chap{run2.m}{\size\begin{verbatim}
(***** RUN2.M *****)

(* handle let/m-exp/s-exp/comments/funny chars/blanks *)
input[x_] := l[m@@ s@@ c@@ Characters@ f@ b@ StringJoin@ x]

(* keep only non-blank printable ASCII codes *)
b[x_] := FromCharacterCode@
 Cases[ ToCharacterCode@ x, n_Integer /; 32 < n < 127 ]

(* handle {dd} chars *)
f[x_] := StringReplace[x,convertmap2]
t[x_] := StringReplace[x,convertmap]
convertmap =
 ( FromCharacterCode@ # -> ToString@ {#-128} )& /@
 Range[128,255]
convertmap2 = convertmap /. (l_->r_)->(r->l)

(* remove comments *)
c["[",x__] := Replace[c@ x,{___,"]",y___}->{y}]
c[x_,y___] := {x} ~Join~ c@ y
c[] := {}

(* handle explicit parens (s-exp) *)
s["(",x__] := Replace[s@ x,{y___,")",z___}->{{y},z}]
s[x_,y___] := {x} ~Join~ s@ y
s[] := {}

(* handle implicit parens (m-exp) *)
get[c_,i_,x_] := {{c}~Join~Take[x,i]} ~Join~ Drop[x,i]
m[c:("+"|"-"|"."|"'"|","|"!"),x__] := get[c,1,m@ x]
m[c:("*"|"="|"&"|"?"),x__] := get[c,2,m@ x]
m[c:("/"|":"),x__] := get[c,3,m@ x]
m[")",y___] := {{}} ~Join~ m@ y
m["\"",")",y___] := {{}} ~Join~ m@ y
m["\"",x_,y___] := {x} ~Join~ m@ y
m[{x___},y___] := {m@ x} ~Join~ m@ y
m[x_,y___] := {x} ~Join~ m@ y
m[] := {}

(* handle definitions (let) *)
l[x_] := x //. {":",{func_,vars___},def_,body_} ->
 {{"'",{"&",{func},body}},{"'",{"&",{vars},def}}} \
 //. {":",var_,def_,body_} ->
 {{"'",{"&",{var},body}},def}

(* output S-exp *)
output[x_String] := x
output[{x___}] := StringJoin["(", output /@ {x}, ")"]

blanks = StringJoin@ Table[" ",{12}]

print[x_,y_] := print1[t@ x,t@ y]
print1[x_,y_] := (print2[x,StringTake[y,50]];
 print1["",StringDrop[y,50]]) /; StringLength[y] > 50
print1[x_,y_] := print2[x,y]
print2[x_,y_] := print3[StringTake[x<>blanks,12]<>y]
print3[x_] := (Print[x]; WriteString[o,x,"\n"])

wrap[e_] :=
 If[ names === {}, e, {{"'",{"&",names,e}}} ~Join~ defs ]

let[n_,d_] :=
(
 print[ output@ n<> ":", output@ d ];
 names = {n} ~Join~ names;
 defs = {{"'",d}} ~Join~ defs;
)

run[fn_,whoami_,outputsuffix_] :=
(
 names = defs = {};
 t0 = SessionTime[];
 o = OpenWrite[fn<>outputsuffix];
 print3["Start of "<>whoami<>" run of "<>fn<>".lisp"];
(
 print3@ "";
 clock = 0;
 Replace[#,{
 {"&",{func_,vars___},def_} :> let[func,{"&",{vars},def}],
 {"&",var_,def_} :> let[var,eval@ def],
 _ :> print[ "value", output@ eval@ #]
 }];
 If[clock != 0, print[ "cycles", ToString@ clock ]];
)& /@ (input@ ReadList[fn<>".lisp",Record]);
 print3@ "";
 print3@ StringForm[
 "Elapsed time `` seconds",
 Round[SessionTime[]-t0]
 ];
 Close@ o
)

runall := run /@ {"test","eval","eval2","eval3","omega"}

$RecursionLimit = $IterationLimit = Infinity
SetOptions[$Output,PageWidth->63];
\end{verbatim}
}\chap{xpnd.m}{\size\begin{verbatim}
(***** XPND.M *****)

Off[ General::spell, General::spell1 ]

run[fn_String] := Module[ {p, o},

(* program p is list of instructions of form: l, op[r,s], *)
p = Get[fn<>".rm"];

SetOptions[$Output,PageWidth->62];
Format[LineBreak[_]] = "";
Format[Continuation[_]] = " ";
Print@ "(**** before ****)";
Print@ Short[InputForm@p,10];

p = p //. {
set[x_,x_] ->
 {},
split[h_,t_,s_] ->
 {set[source,s], jump[linkreg3,split$routine],
 set[h,target], set[t,target2]},
hd[t_,s_] ->
 split[t,target2,s],
tl[t_,s_] ->
 split[target,t,s],
empty[r_] ->
 {set[r,")"], left[r,"("]},
atom[r_,l_] ->
 {neq[r,"(",l], set[work,r], right[work], eq[work,")",l]},
jn[i_,x_,y_] ->
 {set[source,x], set[source2,y], jump[linkreg3,jn$routine],
 set[i,target]},
push[x_] ->
 {set[source,x], jump[linkreg2,push$routine]},
pop[x_] ->
 {jump[linkreg2,pop$routine], set[x,target]},
popl[x_,y_] ->
 split[x,y,y]
};

p = Flatten@ p;

p = p /. op_[l___, x_String, r___]
 :> ( ToExpression[ ToString@ op<> "i" ] )[l,x,r];

p = p //. {l___, x_Symbol, y_, r___}
 -> {l, label[x,y], r};

labels =
 ( ToExpression[ "l"<> ToString@ # ] )& /@ Range@ Length@ p;

p = MapThread[ Replace[#1,
 {label[x__] -> label[x], x_ -> label[#2,x]} ]&,
 {p,labels} ];

p = p /. label[x_,op_[y___]] -> {x,op,y};

r[x_] := ToExpression["reg$"<> ToString@ x]; (* register *)
l[x_] := ToExpression["lab$"<> ToString@ x]; (* label *)
i[x_] := x; (* immediate field *)

t[x_] := x /. {
 {a_,op:halt|dump} :> {l@ a, op},
 {a_,op:goto,b_} :> {l@ a, op, l@ b},
 {a_,op:jump,b_,c_} :> {l@ a, op, r@ b, l@ c},
 {a_,op:goback|right|out,b_} :> {l@ a, op, r@ b},
 {a_,op:eq|neq,b_,c_,d_} :> {l@ a, op, r@ b, r@ c, l@ d},
 {a_,op:eqi|neqi,b_,c_,d_} :> {l@ a, op, r@ b, i@ c, l@ d},
 {a_,op:left|set,b_,c_} :> {l@ a, op, r@ b, r@ c},
 {a_,op:lefti|seti,b_,c_} :> {l@ a, op, r@ b, i@ c} };

p = t /@ p;

Print@ "(**** after ****)";
Print@ Short[InputForm@p,10];

o = OpenWrite[fn<>".xrm",PageWidth->62];
Write[o,p];
Close@ o

]

runall := run /@ {"example","test","lisp"}
\end{verbatim}
}\chap{rm2c.m}{\size\begin{verbatim}
(* RM2C.M *)

p = <<lisp.xrm
p = (ToString /@ #)& /@ p
p = p /. {"'" -> "\\'", "\0" -> "\\0"}
labels = #[[1]]& /@ p
Evaluate[ next /@ labels ] = RotateLeft@ labels
registers =
 Select[ Union@ Flatten@ p, StringMatchQ[#,"reg$*"]& ]

o = OpenWrite@ "lispm.c"
put[x_] := WriteString[o,StringReplace[x,map],"\n"]

map = {}

put@ "/* LISP interpreter running on register machine */"
put@ "#include <stdio.h>"
put@ "#define size 100000"
put@ ""
put@ "main() /* lisp main program */"
put@ "{"
put@ "static char *label[] = {"
(
 map = {"R" -> #};
 put@ "\"(R)\","
)& /@ labels
put@ "\"\"}; /* end of label table */"
put@ ""
put@ "char c, *i, *j, *k;"
put@ "long n;"
put@ "double cycles = 0.0;"
put@ ""
(
 map = "R" -> #;
 put@ "char $R[size] = \"\", *R = $R;"
)& /@ registers
put@ ""
put@ "while ((c = getchar()) != '\\n') *++reg$expression = c;"
put@ ""

Cases[p,
 {l_,op_,a_:"",b_:"",c_:""} :>
 (map =
 {
 "L" -> l, "O" -> op, "A" -> a, "B" -> b, "C" -> c,
 "N" -> StringReverse@ next@ l
 };
 put@ ("/* L: O A,B,C */");
 put@ "L: cycles += 1.0;";
 put@ Switch[
 ToExpression@op,
 dump, "/* not supported */",
 halt, "goto termination_routine;",
 goto, "goto A;",
 goback, "k = A;\ngoto goback_routine;",
 eqi, "if (*A == 'B') goto C;",
 neqi, "if (*A != 'B') goto C;",
 eq, "if (*A == *B) goto C;",
 neq, "if (*A != *B) goto C;",
 right, "if (A != $A) --A;",
 lefti,
 "if (A == ($A+size)) goto storage_full;"
 ~StringJoin~ "\n*++A = 'B';",
 left,
 "if (A == ($A+size)) goto storage_full;"
 ~StringJoin~ "\n*++A = *B;\nif (B != $B) --B;",
 seti,
 If[ b === "\\0",
 "A = $A;",
 "*(A = ($A+1)) = 'B';"
 ],
 set,
 "A = $A;\ni = $B;\nwhile (i < B) *++A = *++i;",
 out,
 "i = $A;\nwhile (i < A) putchar(*++i);\nputchar('\\n');",
 jump,
 "A = $A;\ni = \")N(\";\nwhile ((*++A = *i++) != '(');"
 ~StringJoin~ "\ngoto B;"
 ]
 )
]

put@ ""
put@ ("goto termination_routine; " ~StringJoin~
 "/* in case fell through without halting */")
put@ ""
put@ "goback_routine: n = 0;\n"
put@ "bump_label: i = k;\nj = label[n++];"
put@ "while (*j != '\\0') if (*i-- != *j++) goto bump_label;"
put@ ""
put@ "switch (n) {"
MapThread[
 (
 map = {"L" -> #1, "I" -> #2};
 put@ "case I: goto L;"
 )&,
 {labels,ToString /@ Range[1,Length@labels]}
]

put@ "default:"
put@ "printf(\"!retsasid kcabog\");\ngoto finish;"
put@ "} /* end of switch */"
put@ ""
put@ "storage_full:"
put@ "printf(\"!lluf egarots\");"
put@ "goto finish;"
put@ ""
put@ "termination_routine:"
put@ "i = $reg$value;"
put@ "while (i < reg$value) putchar(*++i);"
put@ "finish:"
put@ "printf(\"\\n%.0f\\n\",cycles);"
put@ ""
put@ "} /* end of lisp machine! */"

Close@ o

(* compile resulting C program *)
Print@ "!cc -O -olispm lispm.c"
!cc -O -olispm lispm.c
\end{verbatim}
}\chap{lisp.c}{\size\begin{verbatim}
/* high speed LISP interpreter */

#include <stdio.h>

#define SIZE 10000000 /* numbers of nodes of tree storage */
#define LAST_ATOM 255 /* highest integer value of character */
#define nil 0 /* null pointer in tree storage */

long hd[SIZE], tl[SIZE]; /* tree storage */
long vlst[LAST_ATOM]; /* bindings of each atom */
long next = LAST_ATOM+1; /* next free cell in tree storage */

void initialize_atoms(void); /* initialize atoms */
void clean_env(void); /* clean environment */
void restore_env(void); /* restore dirty environment */
long eval(long e, long d); /* evaluate expression */
/* evaluate list of expressions */
long evalst(long e, long d);
/* bind values of arguments to formal parameters */
void bind(long vars, long args);
long at(long x); /* atomic predicate */
long jn(long x, long y); /* join head to tail */
long eq(long x, long y); /* equal predicate */
long cardinality(long x); /* number of elements in list */
long out(long x); /* output expression */
void out2(long x); /* really output expression */
long in(); /* input expression */

main() /* lisp main program */
{
 long d = 999999999; /* "infinite" depth limit */
 initialize_atoms();
 /* read in expression, evaluate it, & write out value */
 out(eval(in(),d));
}

void initialize_atoms(void) /* initialize atoms */
{
 long i;
 for (i = 0; i <= LAST_ATOM; ++i) {
 hd[i] = tl[i] = i; /* so that hd & tl of atom = atom */
 /* initially each atom evaluates to self */
 vlst[i] = jn(i,nil);
 }
}

long jn(long x, long y) /* join two lists */
{
 /* if y is not a list, then jn is x */
 if ( y != nil && at(y) ) return x;

 if (next > SIZE) {
 printf("Storage overflow!\n");
 exit(0);
 }

 hd[next] = x;
 tl[next] = y;

 return next++;
}

long at(long x) /* atom predicate */
{
 return ( x <= LAST_ATOM );
}

long eq(long x, long y) /* equal predicate */
{
 if (x == y) return 1;
 if (at(x)) return 0;
 if (at(y)) return 0;
 if (eq(hd[x],hd[y])) return eq(tl[x],tl[y]);
 return 0;
}

long eval(long e, long d) /* evaluate expression */
{
/*
 e is expression to be evaluated
 d is permitted depth - integer, not pointer to tree storage
*/
 long f, v, args, x, y, vars, body;

 /* find current binding of atomic expression */
 if (at(e)) return hd[vlst[e]];

 f = eval(hd[e],d); /* evaluate function */
 e = tl[e]; /* remove function from list of arguments */
 if (f == ')') return ')'; /* function = error value? */

 if (f == '\'') return hd[e]; /* quote */

 if (f == '/') { /* if then else */
 v = eval(hd[e],d);
 e = tl[e];
 if (v == ')') return ')'; /* error? */
 if (v == '0') e = tl[e];
 return eval(hd[e],d);
 }

 args = evalst(e,d); /* evaluate list of arguments */
 if (args == ')') return ')'; /* error? */

 x = hd[args]; /* pick up first argument */
 y = hd[tl[args]]; /* pick up second argument */

 switch (f) {
 case '+': return hd[x];
 case '-': return tl[x];
 case '.': return (at(x) ? '1' : '0');
 case ',': return out(x);
 case '=': return (eq(x,y) ? '1' : '0');
 case '*': return jn(x,y);
 }

 if (d == 0) return ')'; /* depth exceeded -> error! */
 d--; /* decrement depth */

 if (f == '!') {
 clean_env(); /* clean environment */
 v = eval(x,d);
 restore_env(); /* restore unclean environment */
 return v;
 }

 if (f == '?') {
 x = cardinality(x); /* convert s-exp into number */
 clean_env();
 v = eval(y,(d <= x ? d : x));
 restore_env();
 if (v == ')') return (d <= x ? ')' : '?');
 return jn(v,nil);
 }

 f = tl[f];
 vars = hd[f];
 f = tl[f];
 body = hd[f];

 bind(vars,args);

 v = eval(body,d);

 /* unbind */
 while (at(vars) == 0) {
 if (at(hd[vars]))
 vlst[hd[vars]] = tl[vlst[hd[vars]]];
 vars = tl[vars];
 }

 return v;
}

void clean_env(void) /* clean environment */
{
 long i;
 for (i = 0; i <= LAST_ATOM; ++i)
 vlst[i] = jn(i,vlst[i]); /* clean environment */
}

void restore_env(void) /* restore unclean environment */
{
 long i;
 for (i = 0; i <= LAST_ATOM; ++i)
 vlst[i] = tl[vlst[i]]; /* restore unclean environment */
}

long cardinality(long x) /* number of elements in list */
{
 if (at(x)) return 0;
 return 1+cardinality(tl[x]);
}

/* bind values of arguments to formal parameters */
void bind(long vars, long args)
{
 if (at(vars)) return;
 bind(tl[vars],tl[args]);
 if (at(hd[vars]))
 vlst[hd[vars]] = jn(hd[args],vlst[hd[vars]]);
}

long evalst(long e, long d) /* evaluate list of expressions */
{
 long x, y;
 if (at(e)) return nil;
 x = eval(hd[e],d);
 if (x == ')') return ')';
 y = evalst(tl[e],d);
 if (y == ')') return ')';
 return jn(x,y);
}

long out(long x) /* output expression */
{
 out2(x);
 putchar('\n');
 return x;
}

void out2(long x) /* really output expression */
{
 if ( at(x) && x != nil ) {putchar(x); return;}
 putchar('(');

 while (at(x) == 0) {
 out2(hd[x]);
 x = tl[x];
 }

 putchar(')');
}

long in() /* input expression */
{
 long c = getchar(), first, last, next;
 if (c != '(') return c;
 /* list */
 first = last = jn(nil,nil);
 while ((next = in()) != ')')
 last = tl[last] = jn(next,nil);
 return tl[first];
}
\end{verbatim}
}\chap{test.lisp}{\size\begin{verbatim}
[ LISP test run ]
'(abc)
+'(abc)
-'(abc)
*'(ab)'(cd)
.'a
.'(abc)
='(ab)'(ab)
='(ab)'(ac)
-,-,-,-,-,-,'(abcdef)
/0'x'y
/1'x'y
!,'/1'x'y
(*"&*()*,'/1'x'y())
('&(xy)y 'a 'b)
: x 'a : y 'b *x*y()
[ first atom ]
: (Fx)/.,xx(F+x) (F'((((a)b)c)d))
[ concatenation ]
:(Cxy) /.,xy *+x(C-xy) (C'(ab)'(cd))
?'()'
:(Cxy) /.,xy *+x(C-xy) (C'(ab)'(cd))
?'(1)'
:(Cxy) /.,xy *+x(C-xy) (C'(ab)'(cd))
?'(11)'
:(Cxy) /.,xy *+x(C-xy) (C'(ab)'(cd))
?'(111)'
:(Cxy) /.,xy *+x(C-xy) (C'(ab)'(cd))
?'(1111)'
:(Cxy) /.,xy *+x(C-xy) (C'(ab)'(cd))
[ d: x goes to (xx) ]
& (dx) *,x*x()
[ e really doubles length of string each time ]
& (ex) *,xx
(d(d(d(d(d(d(d(d()))))))))
(e(e(e(e(e(e(e(e()))))))))
\end{verbatim}
}\chap{eval.lisp}{\size\begin{verbatim}

[[[ LISP semantics defined in LISP ]]]

[ (Vse) = value of S-expression s in environment e.
  If a new environment is created it is displayed. ]
& (Vse)
  /.s /.es /=s+e+-e (Vs--e)
  ('&(f) [ f is the function ]
   /=f"' +-s
   /=f". .(V+-se)
   /=f"+ +(V+-se)
   /=f"- -(V+-se)
   /=f", ,(V+-se)
   /=f"= =(V+-se)(V+--se)
   /=f"* *(V+-se)(V+--se)
   /=f"/ /(V+-se)(V+--se)(V+---se)
     (V+--f,(N+-f-se)) [ display new environment ]
 (V+se)) [ evaluate function f ]

[ (Nxae) = new environment created from list of
  variables x, list of unevaluated arguments a, and
  previous environment e. ]
& (Nxae) /.xe *+x*(V+ae)(N-x-ae)

[ Test function (Fx) = first atom in the S-expression x. ]
& (Fx)/.xx(F+x)        [ end of definitions ]

(F'(((ab)c)d))           [ direct evaluation ]

(V'(F'(((ab)c)d))*'F*F()) [ same thing but using V ]

\end{verbatim}
}\chap{eval2.lisp}{\size\begin{verbatim}

[[[ Normal LISP semantics defined in "Sub-Atomic" LISP ]]]

[ (Vse) = value of S-expression s in environment e.
  If a new environment is created it is displayed. ]
& (Vse)
  /.+s         /=s+e+-e (Vs--e)
  /=+s'(QUOTE) +-s
  /=+s'(ATOM)  /.+(V+-se)'(T)'(NIL)
  /=+s'(CAR)   +(V+-se)
  /=+s'(CDR)   : x -(V+-se) /.x'(NIL)x
  /=+s'(OUT)   ,(V+-se)
  /=+s'(EQ)    /=(V+-se)(V+--se)'(T)'(NIL)
  /=+s'(CONS)  : x (V+-se) : y (V+--se) /=y'(NIL) *x() *xy
  /=+s'(COND)  /='(NIL)(V++-se) (V*+s--se) (V+-+-se)
  : f  /.++s(V+se)+s       [ f is ((LAMBDA)((X)(Y))(BODY)) ]
  (V+--f,(N+-f-se))        [ display new environment ]


[ (Nxae) = new environment created from list of
  variables x, list of unevaluated arguments a, and
  previous environment e. ]
& (Nxae) /.xe *+x*(V+ae)(N-x-ae)

[ FIRSTATOM
    ( LAMBDA  ( X )
       ( COND  (( ATOM    X )  X )
               (( QUOTE   T ) ( FIRSTATOM  ( CAR   X )))))
]
& F '
((FIRSTATOM)
    ((LAMBDA) ((X))
       ((COND) (((ATOM)  (X)) (X))
               (((QUOTE) (T)) ((FIRSTATOM) ((CAR) (X))))))
)

[ APPEND
    ( LAMBDA  ( X  Y ) ( COND  (( ATOM   X )  Y )
     (( QUOTE   T ) ( CONS  ( CAR   X )
                            ( APPEND  ( CDR   X )  Y )))))
]
& C '
((APPEND)
    ((LAMBDA) ((X)(Y)) ((COND) (((ATOM) (X)) (Y))
     (((QUOTE) (T)) ((CONS) ((CAR) (X))
                            ((APPEND) ((CDR) (X)) (Y))))))
)

(V'
((FIRSTATOM) ((QUOTE) ((((A)(B))(C))(D))))
F)

(V'
((APPEND) ((QUOTE)((A)(B)(C))) ((QUOTE)((D)(E)(F))))
C)
\end{verbatim}
}\chap{eval3.lisp}{\size\begin{verbatim}

[[[ LISP semantics defined in LISP ]]]
[
  Permissive LISP:
  head & tail of atom = atom,
  join of x with nonzero atom = x,
  initially all atoms evaluate to self,
  only depth exceeded failure!

  (Vsed) =
  value of S-expression s in environment e within depth d.
  If a new environment is created it is displayed.

  d is a natural number which must be decremented
  at each call.  And if it reaches zero, evaluation aborts.
  If depth is exceeded, V returns a special failure value $.
  Evaluation cannot fail any other way!
  Normally, when get value v, if bad will return it as is:
  /=$vv
  To stop unwinding,
  one must convert $ to ? & wrap good v in ()'s.
]
& (Vsed)
  /. s : (Ae) /.e s /=s+e+-e (A--e)
                 [ A is "Assoc" ]
        (Ae)     [ evaluate atom; if not in e, evals to self ]
  : f (V+sed)    [ evaluate the function f ]
  /=$ff          [ if evaluation of function failed, give up ]
  /=f"' +-s      [ do "quote" ]
  /=f"/ : p (V+-sed) /=$pp /=0p (V+---sed) (V+--sed)
                 [ do "if" ]
  : (Wl) /.ll : x (V+led) /=$xx : y (W-l) /=$yy *xy
                 [ W is "Evalst" ]
  : a (W-s)      [ a is the list of argument values ]
  /=$aa          [ evaluation of arguments failed, give up ]
  : x +a         [ pick up first argument ]
  : y +-a        [ pick up second argument ]
  /=f". .x       [ do "atom" ]
  /=f"+ +x       [ do "head" ]
  /=f"- -x       [ do "tail" ]
  /=f", ,x       [ do "out"  ]
  /=f"= =xy      [ do "eq"   ]
  /=f"* *xy      [ do "join" ]
  /.d   $        [ fail if depth already zero ]
  : d   -d       [ decrement depth ]
  /=f"! (Vx()d)  [ do "eval"; use fresh environment ]
  /=f"?          [ do "depth-limited eval" ]
     : (Lij) /.i1 /.j0 (L-i-j)
                 [ natural # i is less than or equal to j ]
     /(Ldx) : v (Vy()d) /=$vv *v()
                 [ old depth more limiting; keep unwinding ]
            : v (Vy()x) /=$v"? *v()
                 [ new depth limit more limiting;
                   stop unwinding ]
                 [ do function definition ]
  : (Bxa) /.xe *+x*+a(B-x-a)
                 [ B is "Bind" ]
  (V+--f,(B+-fa)d) [ display new environment ]

[ Test function (Cxy) = concatenate list x and list y. ]

[ Define environment for concatenation. ]
& E '( C &(xy) /.xy *+x(C-xy) )
(V '(C'(ab)'(cd)) E '())
(V '(C'(ab)'(cd)) E '(1))
(V '(C'(ab)'(cd)) E '(11))
(V '(C'(ab)'(cd)) E '(111))
\end{verbatim}
}\chap{omega.lisp}{\size\begin{verbatim}
[
  Make a list of strings into a prefix-free set
  by removing duplicates.  Last occurrence is kept.
]
& (Rx)
[ P-equiv: are two bit strings prefixes of each other ? ]
: (Pxy) /.x1 /.y1 /=+x+y (P-x-y) 0
[ is x P-equivalent to a member of l ? ]
: (Mxl) /.l0 /(Px+l) 1 (Mx-l)
[ body of R follows: ]
/.xx : r (R-x) /(M+xr) r *+xr

[
  K th approximation to Omega for given U.
]
& (WK)
: (Cxy) /.xy *+x(C-xy)           [ concatenation (set union) ]
: (B)
: k ,(*"&*()*,'k())                    [ write k & its value ]
: s (R(C(Hk)s))   [ add to s programs not P-equiv which halt ]
: s ,(*"&*()*,'s())                    [ write s & its value ]
/=kK (Ms)                [ if k = K, return measure of set s ]
: k *1k                                         [ add 1 to k ]
 (B)
: k ()                                [ initialize k to zero ]
: s ()               [ initialize s to empty set of programs ]
 (B)

[
  Subset of computer programs of size up to k
  which halt within time k when run on U.
]
& (Hk)
[ quote all elements of list ]
: (Qx) /.xx **"'*+x()(Q-x)
[ select elements of x which have property P ]
: (Sx) /.xx /(P+x) *+x(S-x) (S-x)
[ property P
  is that program halts within time k when run on U ]
: (Px) =0.?k(Q*U*x())
[ body of H follows:
  select subset of programs of length up to k ]
(S(Xk))

[
  Produce all bit strings of length less than or equal to k.
  Bigger strings come first.
]
& (Xk)
/.k '(())
: (Zy) /.y '(()) **0+y **1+y (Z-y)
(Z(X-k))

& (Mx)   [ M calculates measure of set of programs ]
[ S = sum of three bits ]
: (Sxyz) =x=yz
[ C = carry of three bits ]
: (Cxyz) /x/y1z/yz0
[ A = addition (left-aligned base-two fractions)
  returns carry followed by sum ]
: (Axy) /.x*0y /.y*0x : z (A-x-y) *(C+x+y+z) *(S+x+y+z) -z
[ M = change bit string to 2**-length of string
  example: (111) has length 3, becomes 2**-3 = (001) ]
: (Mx) /.x'(1) *0(M-x)
[ P = given list of strings,
  form sum of 2**-length of strings ]
: (Px)
   /.x'(0)
   : y (A(M+x)(P-x))
   : z /+y ,'(overflow) 0     [ if carry out, overflow ! ]
   -y                         [ remove carry ]
[ body of definition of measure of a set of programs follows:]
: s (Px)
*+s *". -s                    [ insert binary point ]

[
  If k th bit of string x is 1 then halt, else loop forever.
  Value, if has one, is always 0.
]
& (Oxk) /=0.,k (O-x-k)                                [ else ]
        /.x (Oxk)   [ string too short implies bit = 0, else ]
        /+x 0 (Oxk)

[[[ Universal Computer ]]]

& (Us)

[
  Alphabet:
]
: A '"
((((((((leftparen)(rightparen))(AB))((CD)(EF)))(((GH)(IJ))((
KL)(MN))))((((OP)(QR))((ST)(UV)))(((WX)(YZ))((ab)(cd)))))(((
((ef)(gh))((ij)(kl)))(((mn)(op))((qr)(st))))((((uv)(wx))((yz
)(01)))(((23)(45))((67)(89))))))((((((_+)(-.))((',)(!=)))(((
*&)(?/))((:")(${0}))))(((({1}{2})({3}{4}))(({5}{6})({7}{8}))
)((({9}{10})({11}{12}))(({13}{14})({15}{16})))))((((({17}{18
})({19}{20}))(({21}{22})({23}{24})))((({25}{26})({27}{28}))(
({29}{30})({31}{32}))))(((({33}{34})({35}{36}))(({37}{38})({
39}{40})))((({41}{42})({43}{44}))(({45}{46})({47}{48})))))))
[
  Read 7-bit character from bit string.
  Returns character followed by rest of string.
  Typical result is (A 1111 000).
]
: (Cs)
/.--- ---s (Cs)    [ undefined if less than 7 bits left ]
: (Rx) +-x         [ 1 bit: take right half ]
: (Lx) +x          [ 0 bit: take left half ]
*
 (/+s R L
 (/+-s R L
 (/+--s R L
 (/+---s R L
 (/+----s R L
 (/+-----s R L
 (/+------s R L
  A)))) )))
---- ---s
[
  Read zero or more s-exp's until get to a right parenthesis.
  Returns list of s-exp's followed by rest of string.
  Typical result is ((AB) 1111 000).
]
: (Ls)
: c (Cs)                        [ c = read char from input s ]
/=+c'(right paren) *()-c                       [ end of list ]
: d (Es)                       [ d = read s-exp from input s ]
: e (L-d)                 [ e = read list from rest of input ]
   **+d+e-e                              [ add s-exp to list ]
[
  Read single s-exp.
  Returns s-exp followed by rest of string.
  Typical result is ((AB) 1111 000).
]
: (Es)
: c (Cs)                        [ c = read char from input s ]
/=+c'(right paren) *()-c    [ invalid right paren becomes () ]
/=+c'(left  paren) (L-c)      [ read list from rest of input ]
c                 [ otherwise atom followed by rest of input ]

                    [ end of definitions; body of U follows: ]

: x (Es)   [ split bit string into function followed by data ]
! *+x**"'*-x()()    [ apply unquoted function to quoted data ]


[ Omega ! ]
(W'(1111 111 111))
\end{verbatim}
}\chap{example.rm}{\size\begin{verbatim}
{
 set[b,"\0"],
loop,
 left[b,a],
 neq[a,"\0",loop],
 halt[]
}
\end{verbatim}
}\chap{test.rm}{\size\begin{verbatim}
{
label,
 goto[label],
 jump[c,label],
 goback[c],
 neq[a,"b",label],
 neq[a,b,label],
 eq[a,"b",label],
 eq[a,b,label],
 out[c],
 dump[],
 halt[],
 set[a,"b"],
 set[a,b],
 right[c],
 left[a,"b"],
 left[a,b],
 halt[]
}
\end{verbatim}
}\chap{lisp.rm}{\size\begin{verbatim}
{

(* The LISP Machine! ... *)
(* register machine LISP interpreter *)
(* input in expression, output in value *)

 empty[alist], (* initial association list *)
 set[stack,alist], (* empty stack *)
 set[depth,"_"], (* no depth limit *)
 jump[linkreg,eval], (* evaluate expression *)
 halt[], (* finished ! *)

(* Recursive Return ... *)

returnq,
 set[value,"?"],
 goto[unwind],

return0,
 set[value,"0"],
 goto[unwind],

return1,
 set[value,"1"],

unwind,
 pop[linkreg], (* pop return address *)
 goback[linkreg],

(* Recursive Call ... *)

eval,
 push[linkreg], (* push return address *)
 atom[expression,expression$is$atom],
 goto[expression$isnt$atom],

expression$is$atom,
 set[x,alist], (* copy alist *)
alist$search,
 set[value,expression], (* variable not in alist *)
 atom[x,unwind], (* evaluates to self *)
 popl[y,x], (* pick up variable *)
 popl[value,x], (* pick up its value *)
 eq[expression,y,unwind], (* right one ? *)
 goto[alist$search],

expression$isnt$atom, (* expression is not atom *)
 (* split into function & arguments *)
 split[expression,arguments,expression],
 push[arguments], (* push arguments *)
 jump[linkreg,eval], (* evaluate function *)
 pop[arguments], (* pop arguments *)
 eq[value,")",unwind], (* abort ? *)
 set[function,value], (* remember value of function *)

(* Quote ... *)

 neq[function,"'",not$quote],

 (* ' quote *)
 hd[value,arguments], (* return argument "as is" *)
 goto[unwind],

not$quote,

(* If ... *)

 neq[function,"/",not$if$then$else],

 (* / if *)
 popl[expression,arguments], (* pick up "if" clause *)
 push[arguments], (* remember "then" & "else" clauses *)
 jump[linkreg,eval], (* evaluate predicate *)
 pop[arguments], (* pick up "then" & "else" clauses *)
 eq[value,")",unwind], (* abort ? *)
 neq[value,"0",then$clause], (* predicate considered true *)
 (* if not 0 *)
 tl[arguments,arguments], (* if false, skip "then" clause *)
then$clause, (* pick up "then" or "else" clause *)
 hd[expression,arguments],
 jump[linkreg,eval], (* evaluate it *)
 goto[unwind], (* return value "as is" *)

not$if$then$else,

(* Evaluate Arguments ... *)

 push[function],
 jump[linkreg,evalst],
 pop[function],
 eq[value,")",unwind], (* abort ? *)
 set[arguments,value], (* remember argument values *)
 split[x,y,arguments], (* pick up first argument in x *)
 hd[y,y], (* & second argument in y *)

(* Atom & Equal ... *)

 neq[function,".",not$atom],

 (* . atom *)
 atom[x,return1], (* if argument is atomic return true *)
 goto[return0], (* otherwise return nil *)

not$atom,

 neq[function,"=",not$equal],

(* = equal *)
compare,
 neq[x,y,return0], (* not equal ! *)
 right[x],
 right[y],
 neq[x,"\0",compare],
 goto[return1], (* equal ! *)

not$equal,

(* Head, Tail & Join ... *)

 split[target,target2,x], (* get head & tail of argument *)
 set[value,target],
 eq[function,"+",unwind], (* + pick head *)
 set[value,target2],
 eq[function,"-",unwind], (* - pick tail *)
 jn[value,x,y], (* * join first argument to second argument *)
 eq[function,"*",unwind],

(* Output ... *)

 neq[function,",",not$output],

 (* , output *)
 out[x], (* write argument *)
 set[value,x], (* identity function! *)
 goto[unwind],

not$output,

(* Decrement Depth Limit ... *)

 eq[depth,"_",no$limit],
 set[value,")"],
 atom[depth,unwind], (* if limit exceeded, unwind *)
no$limit,
 push[depth], (* push limit before decrementing it *)
 tl[depth,depth], (* decrement it *)

(* Eval ... *)

 neq[function,"!",not$eval],

 (* ! eval *)
 set[expression,x], (* pick up argument *)
 push[alist], (* push alist *)
 empty[alist], (* fresh environment *)
 jump[linkreg,eval], (* evaluate argument again *)
 pop[alist], (* restore old environment *)
 pop[depth], (* restore old depth limit *)
 goto[unwind],

not$eval,

(* Evald ... *)

 neq[function,"?",not$evald],

 (* ? eval depth limited *)
 set[value,x], (* pick up first argument *)
 set[expression,y], (* pick up second argument *)
 (* First argument of ? is in value and *)
 (* second argument of ? is in expression. *)
 (* First argument is new depth limit and *)
 (* second argument is expression to safely eval. *)
 push[alist], (* save old environment *)
 empty[alist], (* fresh environment *)
 (* decide whether old or new depth restriction is stronger *)
 set[x,depth], (* pick up old depth limit *)
 set[y,value], (* pick up new depth limit *)
 eq[x,"_",new$depth], (* no previous limit, *)
 (* so switch to new one *)
choose,
 atom[x,old$depth], (* old limit smaller, so keep it *)
 atom[y,new$depth], (* new limit smaller, so switch *)
 tl[x,x],
 tl[y,y],
 goto[choose],

new$depth, (* new depth limit more restrictive *)
 set[depth,value], (* pick up new depth limit *)
 neq[depth,"_",depth$okay],
 set[depth,"0"], (* only top level has no depth limit *)
depth$okay,
 jump[linkreg,eval], (* evaluate second argument of ? again *)
 pop[alist], (* restore environment *)
 pop[depth], (* restore depth limit *)
 eq[value,")",returnq], (* convert "no value" to ? *)
wrap,
 empty[source2],
 jn[value,value,source2], (* wrap good value in parentheses *)
 goto[unwind],

old$depth, (* old depth limit more restrictive *)
 jump[linkreg,eval], (* evaluate second argument of ? again *)
 pop[alist], (* restore environment *)
 pop[depth], (* restore depth limit *)
 eq[value,")",unwind], (* if bad value, keep unwinding *)
 goto[wrap], (* wrap good value in parentheses *)

not$evald,

(* Defined Function ... *)

 (* bind *)

 tl[function,function], (* throw away & *)
 (* pick up variables from function definition *)
 popl[variables,function],
 push[alist], (* save environment *)
 jump[linkreg,bind], (* new environment *)
 (* (preserves function) *)

 (* evaluate body *)

 hd[expression,function], (* pick up body of function *)
 jump[linkreg,eval], (* evaluate body *)

 (* unbind *)

 pop[alist], (* restore environment *)
 pop[depth], (* restore depth limit *)
 goto[unwind],

(* Evalst ... *)
(* input in arguments, output in value *)

evalst, (* loop to eval arguments *)
 push[linkreg], (* push return address *)
 set[value,arguments], (* null argument list has *)
 atom[arguments,unwind], (* null list of values *)
 popl[expression,arguments], (* pick up next argument *)
 push[arguments], (* push remaining arguments *)
 jump[linkreg,eval], (* evaluate first argument *)
 pop[arguments], (* pop remaining arguments *)
 eq[value,")",unwind], (* abort ? *)
 push[value], (* push value of first argument *)
 jump[linkreg,evalst], (* evaluate remaining arguments *)
 pop[x], (* pop value of first argument *)
 eq[value,")",unwind], (* abort ? *)
 jn[value,x,value], (* add first value to rest *)
 goto[unwind],

(* Bind ... *)
(* input in variables, arguments, alist, output in alist *)

bind, (* must not ruin function *)
 push[linkreg],
 atom[variables,unwind], (* any variables left to bind? *)
 popl[x,variables], (* pick up variable *)
 push[x], (* save it *)
 popl[x,arguments], (* pick up argument value *)
 push[x], (* save it *)
 jump[linkreg,bind],
 pop[x], (* pop value *)
 jn[alist,x,alist], (* (value alist) *)
 pop[x], (* pop variable *)
 jn[alist,x,alist], (* (variable value alist) *)
 goto[unwind],

(* Push & Pop Stack ... *)

push$routine, (* input in source *)
 jn[stack,source,stack], (* stack = join source to stack *)
 goback[linkreg2],

pop$routine, (* output in target *)
 split[target,stack,stack], (* target = head of stack *)
 goback[linkreg2], (* stack = tail of stack *)

(* Split S-exp into Head & Tail ... *)
(* input in source, output in target & target2 *)

split$routine,
 set[target,source], (* is argument atomic ? *)
 set[target2,source], (* if so, its head & its tail *)
 atom[source,split$exit], (* are just the argument itself *)
 set[target,"\0"],
 set[target2,"\0"],

 right[source], (* skip initial ( of source *)
 set[work,"\0"],
 set[parens,"\0"], (* p = 0 *)

copy$hd,
 neq[source,"(",not$lpar], (* if ( *)
 left[parens,"1"], (* then p = p + 1 *)
not$lpar,
 neq[source,")",not$rpar], (* if ) *)
 right[parens], (* then p = p - 1 *)
not$rpar,
 left[work,source], (* copy head of source *)
 eq[parens,"1",copy$hd], (* continue if p not = 0 *)

reverse$hd,
 left[target,work], (* reverse result into target *)
 neq[work,"\0",reverse$hd],

 set[work,"("], (* initial ( of tail *)
copy$tl,
 left[work,source], (* copy tail of source *)
 neq[source,"\0",copy$tl],

reverse$tl,
 left[target2,work], (* reverse result into target2 *)
 neq[work,"\0",reverse$tl],

split$exit,
 goback[linkreg3], (* return *)

(* Join x & y ... *)

jn$routine, (* input in source & source2, *)
 set[target,source], (* output in target *)
 neq[source2,"(",jn$exit], (* is source2 a list ? *)
 set[target,"\0"], (* if not, join is just source1 *)

 set[work,"\0"],
 left[work,source2], (* copy ( at beginning of source2 *)

copy1,
 left[work,source], (* copy source1 *)
 neq[source,"\0",copy1],

copy2,
 left[work,source2], (* copy rest of source2 *)
 neq[source2,"\0",copy2],

reverse,
 left[target,work], (* reverse result *)
 neq[work,"\0",reverse],

jn$exit,
 goback[linkreg3] (* return *)

}
\end{verbatim}
}\end{document}